\documentclass[aps,prd,groupedaddress]{revtex4-1}
\usepackage{graphicx}
\usepackage{amsmath}
\usepackage{amsfonts}
\usepackage{inputenc}
\usepackage{epsfig}

\begin{document}

\title{Cosmic Expansion via Axion-Induced Quintessence}

\author{M.P. Pierpoint}
\email[Email: ]{M.Pierpoint@lboro.ac.uk}
\affiliation{Department of Physics, Loughborough University, Loughborough, Leicestershire, LE11 3TU, United Kingdom}

\author{F.V. Kusmartsev}
\email[Email: ]{F.Kusmartsev@lboro.ac.uk}
\affiliation{Department of Physics, Loughborough University, Loughborough, Leicestershire, LE11 3TU, United Kingdom}

\date{\today}

\begin{abstract}
   In this paper, dark energy is modelled via a spherically symmetric quintessence scalar field $\varphi$, the dynamics of which are found to be analogous to a pendulum. This is due to a driving \textit{axion} potential $V(\left|\varphi\right|)$, whose origins reside within the study of quantum chromodynamics (QCD). The effect of a cosmological constant $\Lambda$, introduced to represent the vacuum energy of space, is also investigated. Preliminary results suggest that $\Lambda$ is analogous to a spring-constant, and thus determining the elasticity of space. Additionally, a cosmological scale-factor $a(t,r)$, notably with an added spatial dependence, is also considered. We propose that due to this inhomogeneous scale-factor, the energy-density is characteristic of temperature fluctuations observed within the cosmic microwave background. Such fluctuations could ultimately lead to a universe composed of filaments and voids; vast expanses of space, separated by regions of localised matter/energy-density. Finally, we provide a means of screening both the cosmological constant and curvature of the universe, to effective values that are more consistent with experimental observation. 
\end{abstract}


\maketitle

\section{Introduction}

Ever since 1929 when universal expansion was first observed, physicists have wrestled to develop an explanation as to precisely why. From {\it steady-state} universe, to {\it big-bang} and inflationary scenarios; with evermore enticing observational evidence at our disposal, we are beginning to converge upon its precise nature. 

Today's state of cosmology is epitomised by the accelerative expansion of the universe, with the so called {\it dark energy} being responsible for this elusive driving force. Numerous experiments, of which include the WMAP \cite{jarosik} and original COBE \cite{smoot} satellite projects, and further via detailed analyses of Type Ia supernovae \cite{riess,perlmutter}, have proven to exhibit conclusive evidence that our universe is subject to accelerated growth. In particular, both COBE and WMAP have provided detailed maps of the cosmic microwave background (CMB) within the universe. Such patterns are characteristic of the energy-density structure at the instant of matter--radiation decoupling. The conclusion of both studies, was that the principal constituents of the universe take the form of dark matter and dark energy, of which there is believed to be a constant battle between the two \cite{bergstrom,boyle}. One pushes the universe to expand, while the other - to collapse.
More objective observations present a universe whereby billions of stars are self-organised into spiralling galaxies, which group into larger, more stable galactic clusters. These structures extend further into enormous superclusters that thread throughout the universe, all of which are moving apart. These observed structures which are entirely compatible with inflation theory and big-bang, show that an understanding of dark energy and its role to play in structure formation, is one the most fundamental problems in modern cosmology. 

More recently, the Planck satellite sought to map the CMB in greater detail than ever before. The mission which was carried out by the European Space Agency (ESA), certainly held true to its word, and after four years of gathering data, Planck delivered its spectacular results \cite{planck}. A key feature which has further motivated this work, is an apparent lensing of CMB photons due to all intermediary matter. Furthermore, the patterns of localised distortions show no random characteristic, with hot or cold spots moving coherently in a single direction \cite{ucl}. This implies that structure formation in the universe has been influenced by a well-defined external influence. We shall seek to explain how one may obtain such a scenario. 

Einstein's remarkable theory of general relativity, tells us that matter and energy distort space-time. On a cosmic scale, the net matter/energy-density of the universe determines its overall space-time curvature. This, in turn, determines the geometry of the universe (i.e., either open, closed or flat). Most believe that the answers reside within the CMB; the primordial light from some half a million years after the big-bang. The geometry of space affects the observed size of hot and cold spots within the CMB - measurements of these variations have indicated that our observable universe is flat to within an error of one percent \cite{jarosik}. 

As mentioned by Steinhardt \cite{steinhardt}, today's universe is part of an endless cycle of big-bangs and big-crunches, with each cycle lasting the order of a trillion years. Despite the amount of matter and radiation in the universe being reset after each cycle, the cosmological constant $\Lambda$ is not. Instead, this constant gradually diminishes over many cycles to the small value observed today \cite{riess,perlmutter,tegmark,riess1}. Indeed, the proposed cyclic universe can incorporate a dynamical mechanism which automatically relaxes the value of the cosmological constant, via a series of quantum phase transitions. This implies that quantum phenomena are essential to our universe. Conversely, it is unclear how the role of other fields or particles (created via quantum fluctuations, as predicted by QCD) must then also exist, and consequently providing the large value for the vacuum energy \cite{wands,copeland,weinberg}. 

Overall, this motivates us to consider two possible contributions within an {\it effective} cosmological constant $\Lambda_{eff}$ - one originating from a quintessence scalar field, while the other being some constant $\Lambda$, chosen to describe the vacuum energy associated with quantum fluctuations. Within this paper, the scalar field is considered (likewise to \cite{steinhardt,valle}) to be driven by an axion sine-Gordon potential, which shall be elaborated upon in Section II. Ultimately, we would like to show that these two contributions may compensate or screen one another entirely. 
		
\section{An Approach to Unifying Dark Matter and Dark Energy}

\subsection{Dark Matter}

Particle physicists have postulated WIMPs (Weakly Interacting Massive Particles) such as axions, dilatons or neutralinos as dark matter candidates \cite{zioutas,ostriker}, while the nature of dark energy is somewhat more elusive. Heterotic string theory even provides as the candidate, a very light universal axion, convenient to describe the nearly massless pseudoscalar field theory \cite{gaillard}.

There is common belief that dark matter and dark energy have nothing to do with each other. However, it has been shown that both may arise from some kind of scalar field \cite{schunck,mielke1}. Both may account, on different scales, for inflation \cite{mielke2,mielke3}, dark matter halos of galaxies \cite{mielke4,mielke5}, or even dark matter condensations (the so-called boson stars) \cite{mielke7,schunck2003} as candidates for Massive Compact Halo Objects (MACHOs). Independently, the views of superstring theory \cite{liddle} suggest an importance of the scalar field with as small a mass as $\sim 10^{-23}$ eV.

In this paper, we follow an approach similar to those of \cite{steinhardt,valle,mielke6}, where axion-like scalar models with periodic self-interaction have been studied. Additionally, the authors of \cite{fuchs} show that an axion Bose-Einstein condensate can provide a substantial contribution to the observed rotation curves of galaxies \cite{fuchs}, and has probably been observed via gravitational lensing in merging clusters. Recent images captured by the Hubble Space Telescope (HST) reveal a mysterious clump of dark matter, thought to be the remnants of a massive galactic collision \cite{sanders}. It seems that the soliton-type dark matter bullets described in \cite{fuchs} provide a natural explanation as to the formation of such dark matter clumps. 

\subsection{Dark Energy}
\textbf{Cosmological Constant:} At first glance of the Friedmann equations, such a phenomenon can be described by the cosmological constant, for which many sub-candidates have been proposed (cf. \cite{schunck,mielke1} and references within). This was first introduced by Albert Einstein, in order to obtain static, stable solutions to the gravitational field equations. In effect, dark energy was used to prevent the gravitational collapse of the universe. Little was it known at the time, that should spatial inhomogeneities be present post-inflation, these could lead to an unstoppable expansion of the universe. Furthermore, the major crux here is an apparent screening of this parameter; the value predicted by experimental observation \cite{riess,perlmutter,tegmark,riess1} remaining inconsistent with the energy scale predictions from particle physics \cite{wands,copeland,weinberg}. The observed value of $7\times 10^{-30}g/cm^{3}$ (or in natural units $\approx 10^{-35}s^{-2}$), is more than 120 orders of magnitude smaller than the Planck density ($\approx 10^{93}g/cm^{3}$) at the instant of the big-bang \cite{steinhardt}. The value itself is merely representative of an overall averaging of the quantum vacuum fluctuations (the so-called {\it quantum foam}), and thus the characteristic energy-density associated with empty space \cite{wands}. 

\ \newline\textbf{Scalar Fields and Higher Order Curvature Lagrangians:} A scalar field, minimally coupled to Einstein's general relativity is equivalent \cite{schunck,mielke1,benitez,magnano} to a modified gravity in the relativistic framework of higher-order curvature Lagrangians. Such effective Lagrangians may also arise from the low-energy limit of superstrings (cf. for example \cite{carroll}), which use a non-linear higher-order curvature Lagrangian to explain the present cosmic accelerated expansion. Our choice of Lagrangian will be outlined in Section IV.

Scalar fields are something we are very much familiar with. They assign numerical value to all points within the domain for which they exist; the temperature of a room being a prime example. Of all the proposed candidates for dark energy, perhaps the most elegant is the quintessence scalar field $\varphi$. The theory posits that some dynamic function (the scalar field), driven by an inherent universal potential $V(\left|\varphi\right|)$, constitutes the underlying mechanism for the observed expansion of the universe.

\ \newline\textbf{Axions:} The existence of scalar fields is also predicted by the standard model of particle physics and quantum chromodynamics (QCD). However, QCD is afflicted with the issue of strong-CP symmetry breaking. Peccei-Quinn theory seeks to remedy this by adding a CP-violating term (the so-called $\varphi$ parameter) \cite{mielke6,valle} to the Yang-Mills Lagrangian. Not only does the theory predict that $\varphi$ is representative of some dynamical field rather than some constant numerical value, but because quantum fields produce particles, the theory predicts the existence of a new particle also - the axion. This particle, as previously mentioned, is regarded by many as one of the best motivated candidates for cold dark matter (CDM) \cite{carosi}. 

Although the cosmological constant will be able to describe the effects of dark energy, we are curious to consider the contribution of axions - investigating how an induced dynamical scalar field potential may relax this value to within observable parameters. An effective potential $V(\left|\varphi\right|)$, arising from the chiral anomaly after integration of the gluon field, is given as follows \cite{valle,duff},

\begin{equation}
V(\left|\varphi\right|)=\frac{m^{4}}{\lambda}\left[1-\cos\left(\frac{\sqrt{\lambda}}{m}\,\varphi\right)\right]\ .
\label{equation9}
\end{equation}\\
Each of the minima within this potential, are associated with different vacuum states, each possessing the same energy. The curvature of the potential at each minimum is related to the axion mass $m$.

Due to the nature of the potential under consideration, a perfectly apt analogy can be associated with that of a pendulum with a time-dependent radius. Since the attached mass is subject to a gravitational potential energy, a consequent effect will be observed upon the radius when transformed into a kinetic equivalent. As such, the precise motivation of this study has been to investigate how a driven scalar field will influence the cosmological radius (i.e., scale-factor) of the universe.

\section{Hybrid Quintessence}

Into our cosmological recipe, we wish to include all we have touched upon in the previous sections. This includes everything from lower-dimensional metrics, to scalar fields, non-linear potentials, and higher-order curvature Lagrangians (with a dash of Kaluza-Klein).

Since the metric outlines the geometry of the space-time domain which we shall be working in, this needs to be carefully defined. An ideal starting point is the FRW metric for an expanding universe. As previously mentioned, the universe is flat to within an error of one percent \cite{jarosik}, thus we shall assume the curvature $k$ of the universe to be equal to zero. In the familiar (3+1) dimensional universe, the metric in Cartesian coordinates (denoted by $x^{\mu}=\left\{t,x,y,z\right\}$) is given as follows,

\begin{equation}
g_{\mu\nu}=diag[\ -1\ ,\ a(t)^{2}\ ,\ a(t)^{2}\ ,\ a(t)^{2}\ ]\ ,
\label{equation10}
\end{equation}\\
where $diag[\ldots]$ denotes a diagonal matrix. Here we subscribe to the sign convention $(\ -\ ,\ +\ ,\ +\ ,\ +\ )$ of Misner-Thorne-Wheeler (MTW) \cite{misner}. The standard kinetic term of any Lagrangian density is of the form $g^{\mu\nu}\partial_{\mu}\varphi\partial_{\nu}\varphi$, where $\partial_{\mu}\varphi=\partial\varphi/dx^{\mu}=\varphi_{,\mu}$. Note that because $\varphi$ is a scalar quantity, there is no requirement to use the covariant derivative $\nabla_{\mu}\varphi$. Due to the Einstein summation convention, this gives,

\begin{equation}
=-\dot{\varphi}^{2}+a(t)^{-2}\left(\varphi_{,x}^{2}+\varphi_{,y}^{2}+\varphi_{,z}^{2}\right)\ .
\label{equation11}
\end{equation}\\
Assuming a new coordinate $r=\sqrt{x^{2}+y^{2}+z^{2}}$, we can use the chain rule of differentiation to modify the spatial derivatives of (\ref{equation11}) as follows,

\begin{displaymath}
\frac{\partial\varphi}{\partial x}=\frac{\partial\varphi}{\partial r}\frac{\partial r}{\partial x}=\frac{x}{r}\ \varphi_{,r}\ .
\end{displaymath}\\ 
Performing the same transformation for both $\varphi_{,y}$ and $\varphi_{,z}$ gives Eq.(\ref{equation11}) as
\begin{equation}
=-\dot{\varphi}^{2}+a(t)^{-2}\varphi_{,r}^{2}\ .
\label{equation12}
\end{equation}
Thus, our metric can be compactified to a diagonal matrix with only two elements, applicable to a new coordinate system $x^{\mu}=\left\{t,r\right\}$.

Continuing our surgery of the FRW metric, we now seek to add some motivation for including electromagnetic fields. For this, we shall adopt the Kaluza-Klein method, whereby an extra dimension is included. In principle, this alternate dimension (of which we have no experience) may be compactified via a periodic boundary condition to such small size, that it evades even the most powerful particle accelerators. To introduce electromagnetic fields, the vector potential $A_{\mu}$ becomes an integral part of the metric \cite{klein,straub}. Modifying accordingly the metric implied by Eq.(\ref{equation12}) gives,
\begin{equation}
\tilde{g}_{\mu\nu}=\bordermatrix{& & & \cr
	& -1+\xi A_{0}A_{0} & \xi A_{0}A_{1} & \xi A_{0} \cr
	& \xi A_{1}A_{0} & a(t,r)^{2}+\xi A_{1}A_{1} & \xi A_{1} \cr
	& \xi A_{0} & \xi A_{1} & \xi \cr}
	\label{equation13}
\end{equation}\\
where $\xi$ is a constant. This ultimately presents a $(2+1)$ dimensional metric for a new coordinate system $x^{\mu}=\left\{t,r,\chi\right\}$, where $\chi$ is an unseen extra dimension. A tilde is chosen to denote a variable in the Kaluza-Klein framework. However, these can be expressed in terms of the original metric (without vector potential $A_{\mu}$ included). The new metric has three overall effects. Firstly, the determinant of the metric transforms as $\tilde{g}=\det{|\tilde{g}_{\mu\nu}|}\rightarrow \xi g$. Secondly, the Ricci scalar curvature $\tilde{\mathcal{R}}\rightarrow\mathcal{R}+\frac{1}{4}\xi F_{\mu\nu}F^{\mu\nu}$. Finally, the kinetic term $g^{\mu\nu}\partial_{\mu}\varphi\partial_{\nu}\varphi$ becomes gauge invariant, transforming into $\left|\rm{D}\varphi\right|^{2}=(\partial_{\mu}\varphi-i\sqrt{\xi}A_{\mu}\varphi)g^{\mu\nu}(\partial_{\nu}\varphi^{*}+i\sqrt{\xi}A_{\nu}\varphi^{*})$. Here $\varphi^{*}$ may be chosen to denote the complex conjugate of the scalar field. 

Generally speaking, no parameters should depend upon the coordinate $\chi$. Furthermore, it is not often that cosmologists consider a spatially-dependent scale-factor within the universe. We have chosen to include this, in order the model large-scale spatial perturbations. The premise is analogous to that of Einstein standing on a trampoline (the trampoline representing space-time). The presence of Einstein's mass, stretches the material to a larger degree in the immediate vicinity. Towards the edges of the trampoline, little stretching occurs. 

All ingredients have now been thrown into our cosmological soup. All that remains, is to formulate the required Lagrangian density $\mathcal{L}$. For an action $S=\int\mathcal{L}\sqrt{-\xi g}\ d^{3}x$, this gives,

\begin{equation}
S=\int d^{3}x\,\frac{\sqrt{-\xi g}}{2\kappa}\left(\mathcal{R}+2\Lambda+\frac{1}{4}\,\xi F_{\mu\nu}F^{\mu\nu}+\kappa\left[\left|\rm{D}\varphi\right|^{2}-2V(\left|\varphi\right|)\right]\right)
\label{equation14}
\end{equation}\\
where gravitational coupling constant $\kappa=8\pi G$, $\Lambda$ is the cosmological constant, $F_{\mu\nu}=\partial_{\mu}A_{\nu}-\partial_{\nu}A_{\mu}$ is the electromagnetic (field strength) tensor, D$_{\mu}=\partial_{\mu}-i\sqrt{\xi}A_{\mu}$ is the gauge covariant derivative, and $V(\left|\varphi\right|)$ is the driving potential (to be chosen in parallel with Eq.(\ref{equation9})). The definition of $F_{\mu\nu}$ would appear to oblige the use of a covariant derivative (e.g., $\nabla_{\mu}A_{\nu}=\partial_{\mu}\varphi-\Gamma^{\sigma}_{\mu\nu}A_{\sigma}$), since we are working with a non-scalar quantity. However, because the indices of $F_{\mu\nu}$ are anti-symmetric, and the bottom two indices of the Christoffel symbol $\Gamma^{\sigma}_{\mu\nu}$ are symmetric, this therefore reduces to the standard partial derivative.    

Overall, we have a surprisingly beautiful result. Via explicit inclusion of the electromagnetic vector potential $A_{\mu}$ within our metric, we have inadvertently reproduced the Lagrangian term that governs axion electrodynamics $\Delta\mathcal{L}=(\theta e^{2}/4\pi^{2}){\rm tr}F_{\mu\nu}\tilde{F}_{\mu\nu}$ (cf. for details \cite{wilczek,thooft,callan,jackiw}). Here, the parameter $\theta$ denotes a specific vacuum of the potential $V(\varphi)$. 

\section{Screening the Cosmological Constant}

Ultimately, we wish to screen the value of the cosmological constant, to an effective value $\Lambda_{eff}$ which is more consistent with experimental observation. To achieve this, we begin with the original Einstein-Hilbert action given as,

\begin{displaymath}
S=\int d^{4}x\,\frac{\sqrt{-g}}{\kappa}\left(\mathcal{R}_{eff}+2\Lambda_{eff}\right)\ .
\end{displaymath}\\
Comparing with (\ref{equation14}) gives and effective cosmological constant,
\begin{equation}
\Lambda_{eff}=\sqrt{\xi}\,\Lambda+\frac{1}{8}\xi^{3/2}\, F_{\mu\nu}F^{\mu\nu}+\kappa\,\sqrt{\xi}\,\left[\frac{1}{2}\left|\rm{D}\varphi\right|^{2}-V(\left|\varphi\right|)\right]\ ,
\label{equation15}
\end{equation} and an effective scalar curvature,
\begin{displaymath}
\mathcal{R}_{eff}=\sqrt{\xi}\,\mathcal{R}\ .
\end{displaymath} Provided the value of $\xi$ is very small, this may provide a means of screening both values. Upon variation of the action (\ref{equation14}) with respect to the background metric $g^{\mu\nu}$, emerge the Einstein-Maxwell field equations.

\begin{displaymath}
R_{\mu\nu}-\frac{g_{\mu\nu}}{2}\mathcal{R}-g_{\mu\nu}\Lambda=-\frac{\xi}{2}\,F_{\mu\alpha}F^{\alpha}_{\nu}+\frac{\xi}{8}\,g_{\mu\nu}F_{\sigma\alpha}g^{\alpha\beta}F_{\beta\rho}g^{\rho\sigma}
\end{displaymath}
\begin{displaymath}
-\frac{\kappa}{2}\left[(\partial_{\mu}\varphi-iqA_{\mu}\varphi)(\partial_{\nu}\varphi^{*}+iqA_{\nu}\varphi^{*})+\mu\leftrightarrow\nu\right]
\end{displaymath} 
\begin{equation}
+\kappa g_{\mu\nu}\left[\frac{1}{2}\left|\rm{D}\varphi\right|^{2}-V(\left|\varphi\right|)\right]=-\kappa\ T_{\mu\nu}\ ,
\label{equation16}
\end{equation}\\ where we have performed a final symmetrisation process upon the gauge covariant derivative term.

\section{Locally Flat Space-Time}

\begin{figure}[ph]
	\centering
		\includegraphics[width=0.75\textwidth]{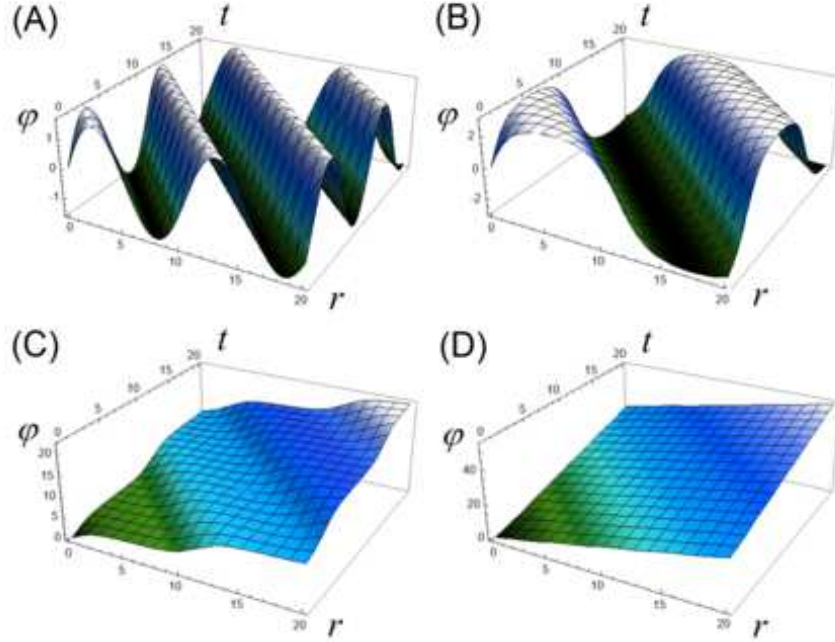}
	\caption{Space-time plots of scalar field $\varphi(t,r)$ for various values of cosmological constant $\Lambda$. Constants $v$, $\lambda$, $m$, and $\kappa$ are all set equal to one. \textbf{(A)} $\Lambda=1$ Here, the system is analogous to a pendulum oscillating back and forth, whilst travelling through the spatial domain with speed $v$. \textbf{(B)} $\Lambda=1.99$ Similar to the previous case, a pendulum continues to oscillate but with much larger amplitude and time period. \textbf{(C)} $\Lambda=2.01$ Here, the scalar field now has enough energy to roll into the next vacuum state, and continues to increase in value. \textbf{(D)} $\Lambda=3$ The system is now analogous to that of an orbiting body performing circular motion.}
	\label{fig1}
\end{figure}

\begin{figure}[ph]
	\centering
		\includegraphics[width=0.75\textwidth]{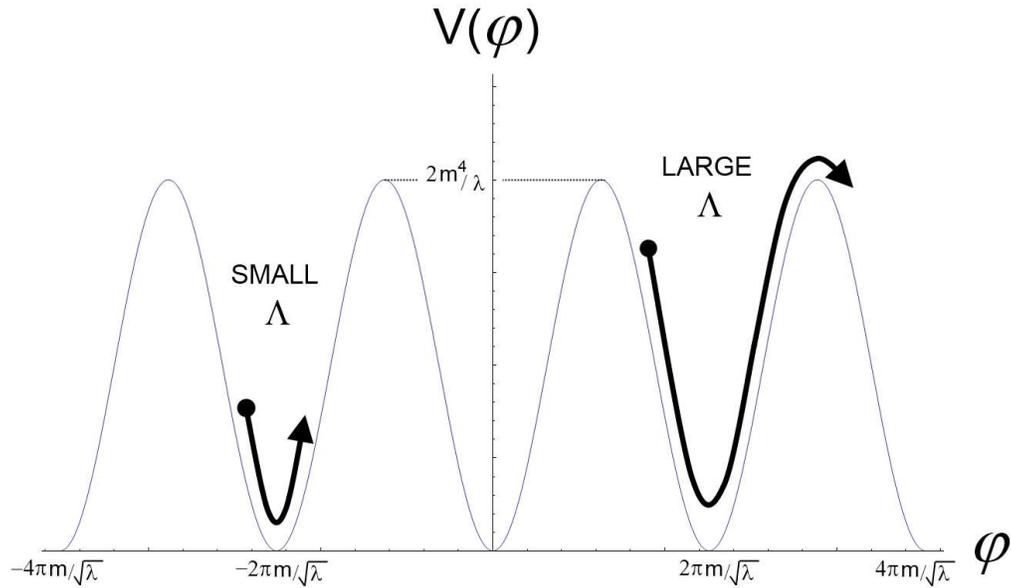}
	\caption{Plotted is the axion potential $V(\varphi)=\frac{m^{4}}{\lambda}\left[1-\cos\left(\frac{\sqrt{\lambda}}{m}\,\varphi\right)\right]$. Additionally emphasised is the effect of the cosmological constant $\Lambda$. For small values of $\Lambda$, the scalar field will \textit{roll} inside one of the vacuum states (cf. FIG. 7.1\textbf{(A)}). These vacuum states are centered about the positions $\varphi=2\pi nm/\sqrt{\lambda}$, for integer values of $n$. Conversely, for large values of $\Lambda$, the scalar field has the capability to roll from one vacuum state and into another (cf. FIGS. 7.1\textbf{(C-D)}).}
	\label{fig2}
\end{figure}

We first consider the simplest case of a locally flat space-time, subject to the Minkowski metric $\eta_{\mu\nu}=diag(-1,1)$. For this, we shall adopt the Hamiltonian formalism, rather than the Lagrangian equivalent. The Hamiltonian density of the system will then be chosen, as necessary, to be proportional to the cosmological constant $\Lambda$. From Eq.(\ref{equation14}), we find that this this should be equal to $\Lambda/\kappa$. The potential chosen to drive the scalar field is the axion potential (\ref{equation9}). For simplicity, we shall assume no electromagnetic interaction $A_{\mu}=0$, and a real scalar field $\varphi=\varphi^{*}$ which does not depend upon our holographic dimension $\chi$. \newline\newline The Hamiltonian density is now read as;

\begin{displaymath}
\mathcal{H}=\frac{\dot{\varphi}^{2}}{2}+V(\varphi,\varphi_{r})\ ,
\end{displaymath}

\begin{equation}
\Longrightarrow\ \ \ \frac{\Lambda}{\kappa}=\frac{\dot{\varphi}^{2}}{2}+\frac{\varphi_{r}^{2}}{2}+\frac{m^{4}}{\lambda}\left[1-\cos\left(\frac{\sqrt{\lambda}}{m}\,\varphi\right)\right]\ .\ \ \ \ \ \ \ \ 
\label{equation24}
\end{equation}\\ As a side remark; for a homogeneous scalar field, Eq.(\ref{equation24}) reduces to the Hamiltonian density for a pendulum. Indeed, the temporal derivative of Eq.(\ref{equation24}) (with $\varphi_{r}=0$), yields the equation of motion for a pendulum. Since much is already known of the simple pendulum as a classical system, we shall not discuss the solutions in detail. However, for an inhomogeneous scalar field $\varphi(t,r)$, Eq.(\ref{equation24}) has the exact solution in terms of the Jacobi amplitude,

\begin{equation}
\varphi(t,r)=\frac{2m}{\lambda}\,{\rm am}\left[\sqrt{\frac{\Lambda\lambda}{2\kappa m^{2}(1+v^{2})}}\,(r+vt)\ \vert\ \frac{2\kappa m^{4}}{\Lambda\lambda}\right]\ .\ \ \ 
\label{equation25}
\end{equation}\\ 
Here we subscribe to the notation adopted by Abramowitz and Stegun - $\varphi=am(u|M)$ where $M=k^{2}$. Plots of this solution can be found in FIG. \ref{fig1} for various choice of $\Lambda$. For convenience, the constants $\lambda$, $m$, $\kappa$ and $v$ are normalised to $1$. The constant $v$ refers to the velocity of the wave, as a fraction of the speed of light (since the speed of light $c=1$).
 FIG. \ref{fig1}\textbf{(A)} is merely analogous to a pendulum swinging back and forth, whilst travelling through the $r$-domain with speed $v$. As we begin to surpass a separatrix value of $\Lambda\approx 2$, the scalar field gains enough angular-momentum to `roll' into the next vacuum state (i.e., minima) of the potential $V(\varphi)$ (cf. FIG. \ref{fig2}). Furthermore, as $\Lambda\rightarrow 3$, the system becomes analogous to an orbiting body performing circular motion.

However, as previously mentioned, the system is partly analogous to that of a swinging pendulum. It was therefore natural to investigate how a pendulum would behave when given an extra degree of freedom (e.g., a time-dependent radius). Rather co-incidentally, the subsequent equations of motion are identical in nature to those for a quintessence scalar field $\varphi$, coupled to a scaling factor $a$ across the spatial component of the metric.

\section{Maximal Spatial Homogeneity: $\varphi(t)$, $a(t)$}
We now consider the maximally homogeneous scenario, with a scalar field $\varphi(t)$ coupled to a scale-factor $a(t)$ across the spatial domain. This scaling factor constitutes the $g_{11}$ component of the flat FRW metric to give $g_{\mu\nu}=diag(-1,a^{2}(t))$. The function is squared to ensure that adjacent points remain a positive distance apart. 

\begin{figure}[ph]
	\centering
		\includegraphics[width=0.75\textwidth]{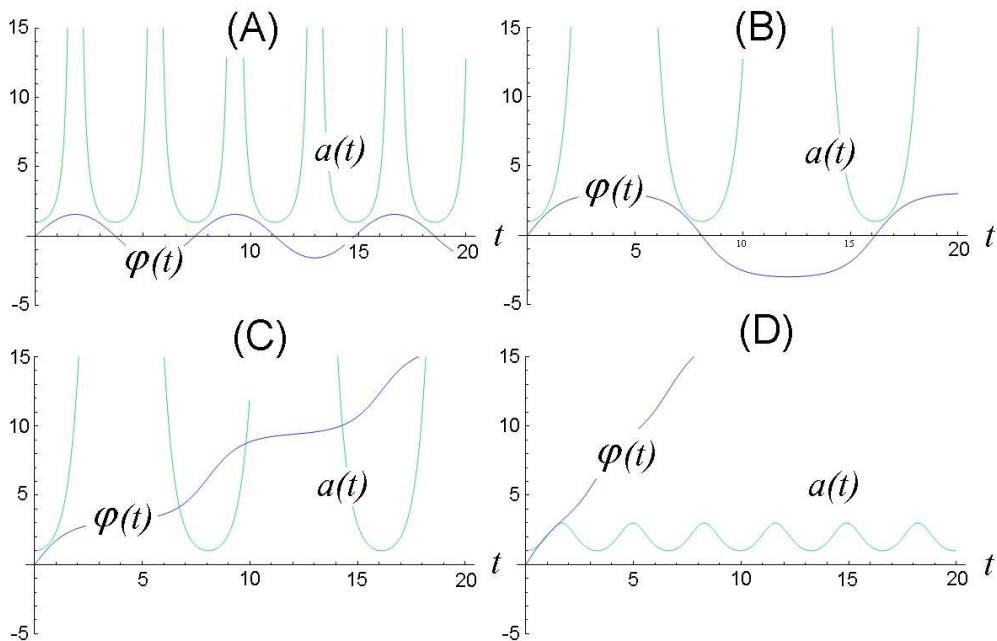}
	\caption{Plots of scalar field (blue) and scale-factor (green) for various choice of $\Lambda$.  Constants $v$, $\lambda$, $m$, and $\kappa$ are all set equal to one. \textbf{(A)} $\Lambda=1$, \textbf{(B)} $\Lambda=1.99$, \textbf{(C)} $\Lambda=2.01$, \textbf{(D)} $\Lambda=3$. For values of $\Lambda>2$, the scalar field continues to increase. As the kinetic $\dot{\varphi}$ contribution diminishes, this energy is transformed into an elastic potential energy, manifesting itself as growth of the scale-factor.}
	\label{fig3}
\end{figure}

All of the necessary parameters are then evaluated, and substituted into the Lagrangian density $\mathcal{L}$ specified within (\ref{equation14}). This is then substituted into the gauge covariant Euler-Lagrange equation \cite{lewis}, 

\begin{equation}
\frac{\partial\left(\sqrt{-g}\mathcal{L}\right)}{\partial\varphi}-\partial_{\mu}\left(\frac{\partial\left(\sqrt{-g}\mathcal{L}\right)}{\partial\left(\partial_{\mu}\varphi\right)}\right)=0\ ,
\label{equation26}
\end{equation}\\ subsequently giving the following two equations of motion:

\begin{equation}
\ddot{\varphi}+\frac{\dot{a}}{a}\dot{\varphi}-\frac{m^{4}}{\lambda}\left[1-\cos\left(\frac{\sqrt{\lambda}}{m}\,\varphi\right)\right]=0\ ,
\label{equation27}
\end{equation}

\begin{equation}
\frac{\Lambda}{\kappa}=\frac{\dot{\varphi}^{2}}{2}+\frac{m^{4}}{\lambda}\left[1-\cos\left(\frac{\sqrt{\lambda}}{m}\,\varphi\right)\right]\ .
\label{equation28}
\end{equation}\newline Eq.(\ref{equation28}) has the following exact solution, 

\begin{equation}
\varphi(t)=\frac{2m}{\lambda}\,{\rm am}\left[\sqrt{\frac{\Lambda\lambda}{2\kappa m^{2}}}\,t\ \vert\ \frac{2\kappa m^{4}}{\Lambda\lambda}\right]\ .
\label{equation28a}
\end{equation}\newline The solution for $\varphi(t)$ given above, is now substituted into Eq.(\ref{equation27}). The subsequent differential equation is then solved for the scale-factor $a(t)$ to give the following Jacobi relation,

\begin{displaymath}
a(t)=\frac{C_{1}}{{\rm dn}\left[\sqrt{\frac{\Lambda\lambda}{2\kappa m^{2}}}\,t\ \vert \frac{2m^{4}\kappa}{\Lambda\lambda}\right]^{2}}\ .
\end{displaymath}\\
The constant $C_{1}$ merely influences the amplitude of $a(t)$. For simplicity, this has been normalised to one. Plots of this solution can be found in FIGS. \ref{fig3}\textbf{(A-D)} for various choice of $\Lambda$.

As previously mentioned, for small values of $\Lambda$, the scalar field does not have the capability to roll from one vacuum state to another (cf. FIG. \ref{fig2}). This implies the scalar field will eventually have a $\dot{\varphi}$ (i.e., kinetic) component equal to zero. However, for a pendulum, as $\dot{\varphi}$ approaches zero, this kinetic energy would transform into an elastic potential energy, and thus an expansion of its radius (i.e., our spatial domain). As the constant $\Lambda$ is increased, so does the overall $\dot{\varphi}$ contribution, and the relative change in scale-factor is found to be much smaller (cf. FIG. \ref{fig3}{\bf (D)}). Therefore, the constant $\Lambda$ can be considered to be analogous to a spring-constant, determining the elasticity of the spatial domain. As previously mentioned, if $\Lambda$ is sufficiently large, the system will exhibit circular motion, and thus the pendulum length (i.e., scale-factor of the universe) will be constant.

\section{Partial Homogeneity}
In this section, we consider an inhomogeneous scale-factor $a(t,r)$, as this possesses a surprisingly simple result. From this point on, we disregard the values of $\Lambda$ which result in an infinite expansion (cf. FIGS. 7.3\textbf{(A-B)}). 

\begin{figure}[p]
	\centering
		\includegraphics[width=0.75\textwidth]{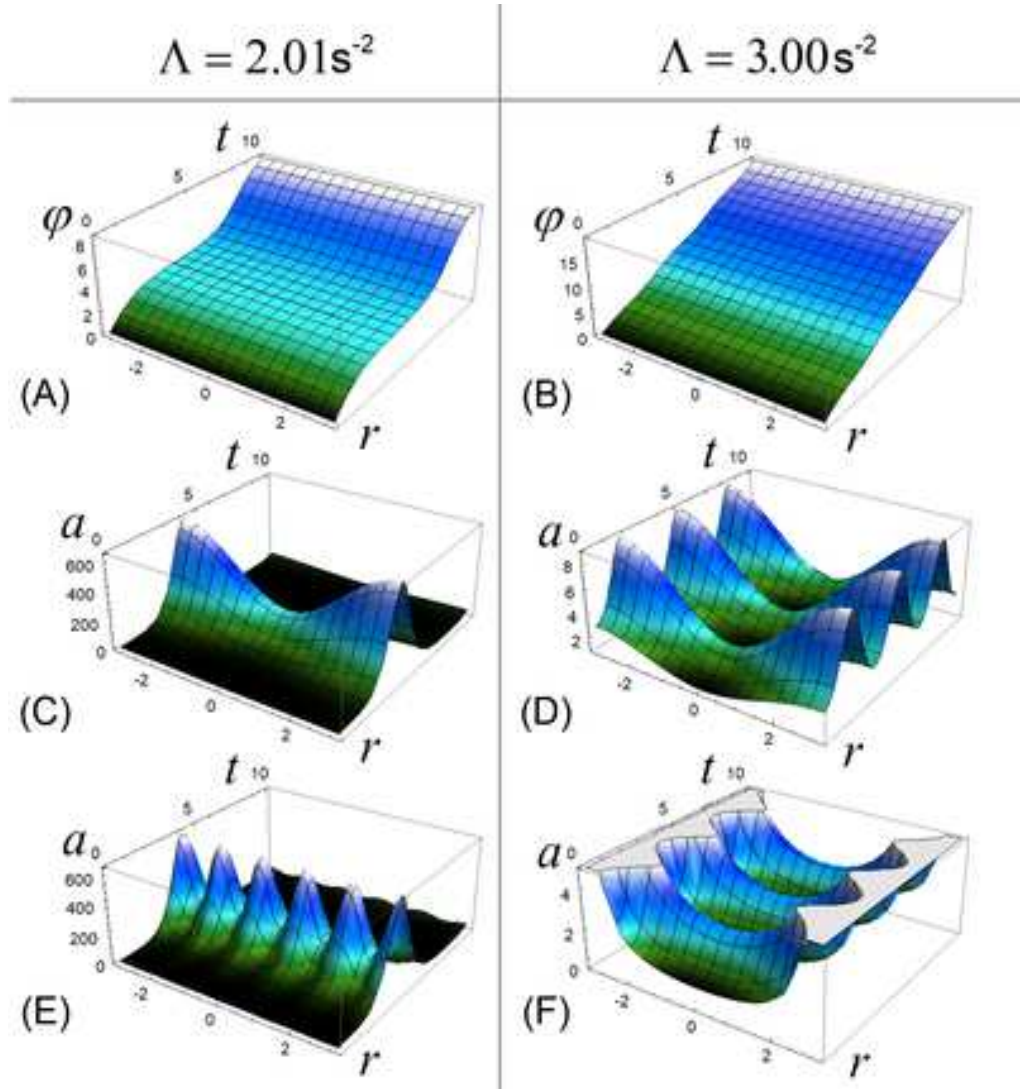}
	\caption{\textbf{(A-B)} Space-Time plots of a homogeneous scalar field $\varphi$ for two differing choices of $\Lambda$. \textbf{(C-D)} Space-Time plots of an inhomogeneous scale-factor with $C_{1}(r)=2-\cos(r)$. \textbf{(E)} Space-Time plot of an inhomogeneous scale-factor with $C_{1}(r)=2-\cos(5r)$. \textbf{(F)} Space-Time plot of an inhomogeneous scale-factor with $C_{1}(r)=\sec(r)$.}
	\label{fig4}
\end{figure}

After performing all the necessary steps, the obtained equations of motion are found to be identical to those for the maximally homogeneous scenario. However, after substituting the solution (\ref{equation28a}) for $\varphi(t)$ into Eq.(\ref{equation27}), we instead solve the differential equation for a scale-factor $a(t,r)$. The solution is given as the following Jacobi relation,

\begin{displaymath}
a(t,r)=\frac{C_{1}(r)}{{\rm dn}\left[\sqrt{\frac{\Lambda\lambda}{2\kappa m^{2}}}\,t\ \vert \frac{2m^{4}\kappa}{\Lambda\lambda}\right]^{2}}\ .
\end{displaymath}\\
This is just the same as before, with the exception that amplitude $C_{1}$ now has an added spatial dependence. This allows one to specify the initial perturbation that is present in the scale-factor. Plots are shown in FIG. \ref{fig4} for various choice of $C_{1}(r)$. Within FIGS. \ref{fig4}\textbf{(C-D)}, we have $C_{1}(r)=2-\cos(r)$. The boundaries of the spatial domain are specified accordingly, so as to satisfy the topological identification $P_{1}=P_{2}\leftrightarrow r(P_{1})=r(P_{2})+2\pi Rn$ where $n\in\mathbb{Z}$. It is evident from the gridlines, that certain regions are subject to a faster rate of expansion. For FIG. \ref{fig4}\textbf{(E)}, five distinct regions across the spatial domain now undergo an accelerated expansion, with the topological identification remaining in-situ. FIG. \ref{fig4}\textbf{(F)} on the other hand, has the effect of modelling an infinite universe (via the scale-factor) within a finite-sized domain. No topological identifications are specified here. 

\begin{figure}[hp]
	\centering
		\includegraphics[width=0.5\textwidth]{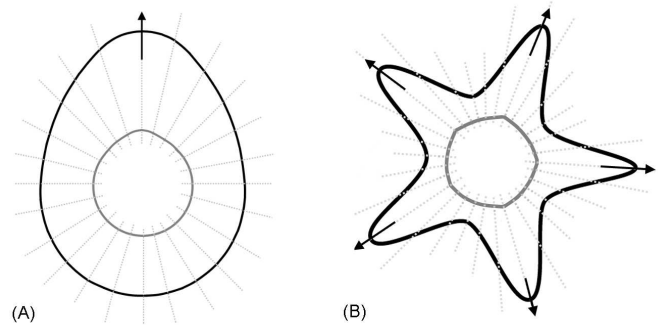}
	\caption{\textbf{(A)} A periodic universe with $C_{1}(r)=2-\cos(r)$. The dashed lines indicate the $r$-coordinate grid axis. The arrow indicates a region with large scale-factor $a(t,r)$. Adjacent points accelerate away from one another with a velocity proportional to their separation; thus mimicking the effect of Hubble's law. \textbf{(B)} A periodic universe with $C_{1}(r)=2-\cos(5r)$. The arrows indicate five distinct regions of accelerated expansion.}
	\label{fig5}
\end{figure}

Physical representations of FIGS. \ref{fig4}\textbf{(C)(E)} are shown in FIGS. \ref{fig5}\textbf{(A-B)} respectively. Regions of large scale-factor have an observable effect upon the $r$-coordinate grid axis. An observer positioned at $r=0$ would observe adjacent $r$-coordinate lines receding at a rate which is proportional to their distance away. This has the desired effect of modelling Hubble's law.
 
\section{Maximum Inhomogeneity\ $\varphi(t,r)$, $a(t,r)$}

\begin{figure}[p]
	\centering
		\includegraphics[width=0.75\textwidth]{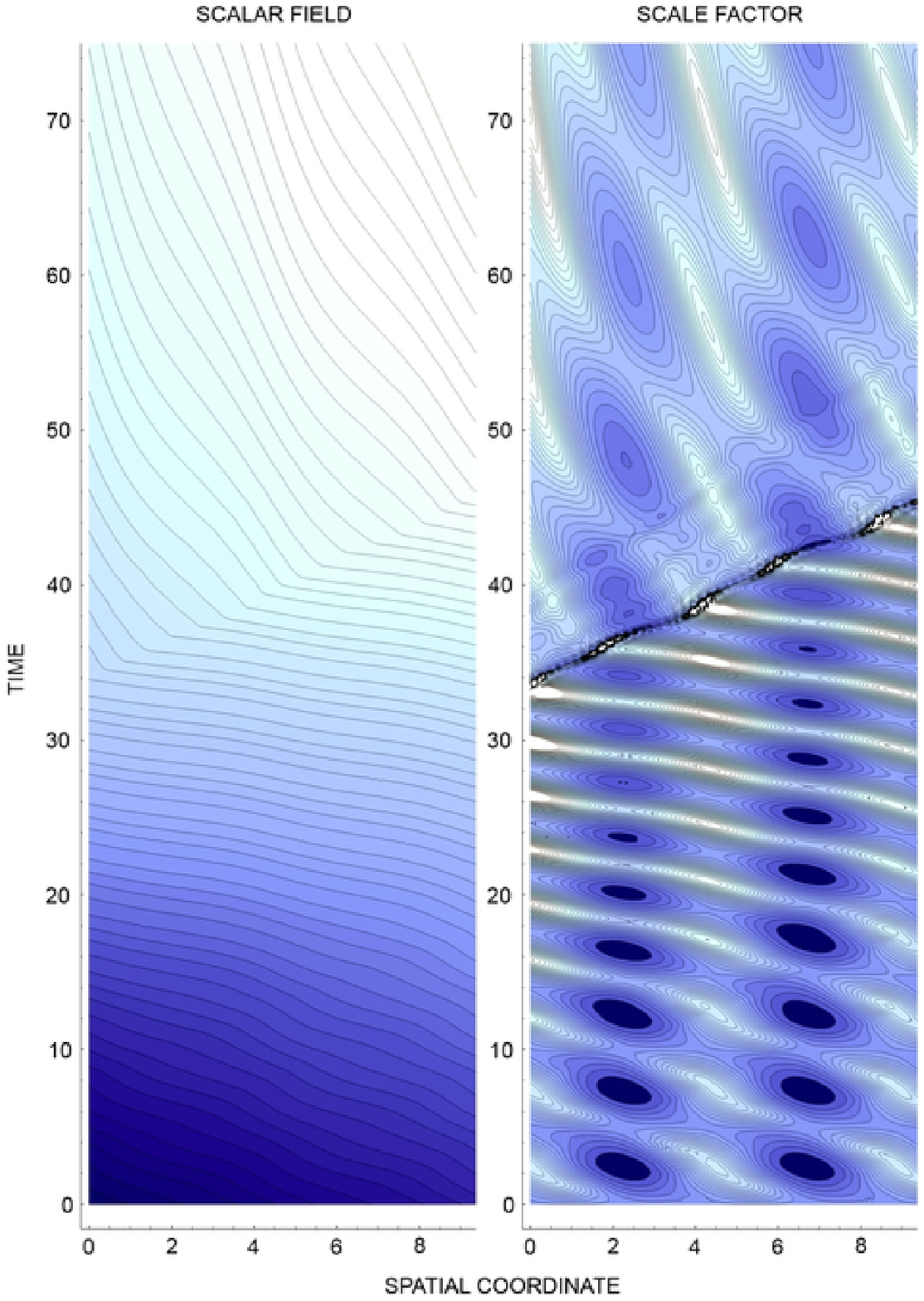}
		\caption{The vertical and horizontal axes represent the temporal and spatial dimensions respectively. The scalar field $\varphi$ (left) is plotted across the range $1.6<\varphi<96.0$ (60 contours), with darker colours representing a smaller value. The scale-factor $a$ (right) is plotted across the range $0.657<a<2.774$ (30 contours), although higher values are omitted.}
		\label{fig20}
		\end{figure}

\begin{figure}[p]
	\centering
		\includegraphics[width=0.75\textwidth]{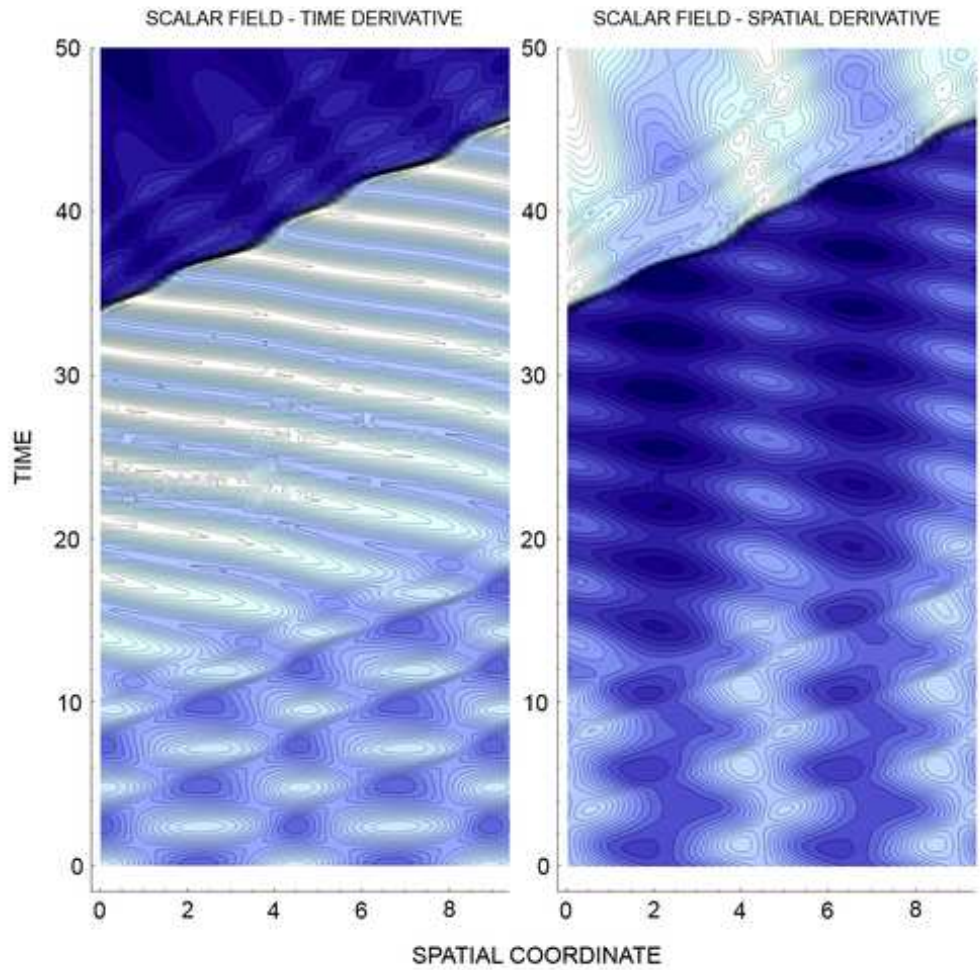}
		\caption{The vertical and horizontal axes represent the temporal and spatial dimensions respectively. The time derivative of scalar field $\dot{\varphi}$ (left) is plotted across the range $0.531<\dot{\varphi}<2.242$ (30 contours), with darker colours representing a smaller value. The spatial gradient of the scalar field $\varphi_{r}$ (right) is plotted across the range $0.5<\varphi_{r}<3.4$ (30 contours).}
		\label{phidot}
		\end{figure}

For this scenario, the derived equations of motion are as follows;

\begin{equation}
a^{3}\ddot{\varphi}-a\varphi_{rr}+a^{2}\,\dot{a}\dot{\varphi}+a_{r}\varphi_{r}-a^{3}\,\frac{m^{3}}{\sqrt{\lambda}}\sin\left({\frac{\sqrt{\lambda}}{m}\,\varphi}\right)=0\ ,
\label{equation29}
\end{equation}

\begin{equation}
\frac{\Lambda}{\kappa}=\frac{\dot{\varphi}^{2}}{2}+\frac{\varphi_{r}^{2}}{2a^{2}}+\frac{m^{4}}{\lambda}\left[1-\cos\left(\frac{\sqrt{\lambda}}{m}\,\varphi\right)\right]\ .
\label{equation30}\end{equation}\\
By specifying initial conditions at $t=t_{0}$ that we have $\varphi=\varphi_0$, $\dot{\varphi}=\dot{\varphi}_{0}$, $\varphi_{r}=\varphi_{0r}$, and $a_{0}=1$, Eq.(\ref{equation30}) can once again be solved analytically;

\begin{equation}
\varphi(t=t_{0},r)=\frac{2m}{\lambda}\,{\rm am}\left[\sqrt{\frac{\Lambda\lambda}{2\kappa m^{2}(1+v^{2})}}\,(r+vt_{0})\ \vert\ \frac{2\kappa m^{4}}{\Lambda\lambda}\right]\ .
\label{equation32}
\end{equation}\newline Furthermore, one may also re-arrange Eq.(\ref{equation30}) for the scale factor $a(t,r)$, and substitute this into Eq.(\ref{equation29}). This then gives an equation of motion that requires solving only in the scalar field $\varphi(t,r)$.

\begin{equation}
a(t,r)=\sqrt{\frac{\kappa\lambda\,\varphi_{r}^{2}}{2\lambda\Lambda-2m^{4}\kappa\left[1-\cos\left(\frac{\sqrt{\lambda}}{m}\,\varphi\right)\right]-\kappa\lambda\dot{\varphi}^{2}}}\ .
\label{equation33}\end{equation}\\  
This can then solved numerically by specifying (\ref{equation32}) and its temporal derivative as the initial conditions. Following the computation of $\varphi(t,r)$, $\dot{\varphi}(t,r)$ and $\varphi_{r}(t,r)$ (cf. FIGS. \ref{fig20} (left) - \ref{phidot} (left/right) respectively), subject to a certain choice of parameters ($\Lambda=3$, $\lambda=1$, $m=1$, $\kappa=1$, $v=0.9$), one may then use Eq.(\ref{equation33}) to retrieve the scale-factor $a(t,r)$. A plot of this can be found in FIG. \ref{fig20} (right). In these figures, the vertical axis corresponds to the time component, and the horizontal axis - the spatial component.

\begin{figure}[p]
	\centering
		\includegraphics[width=0.75\textwidth]{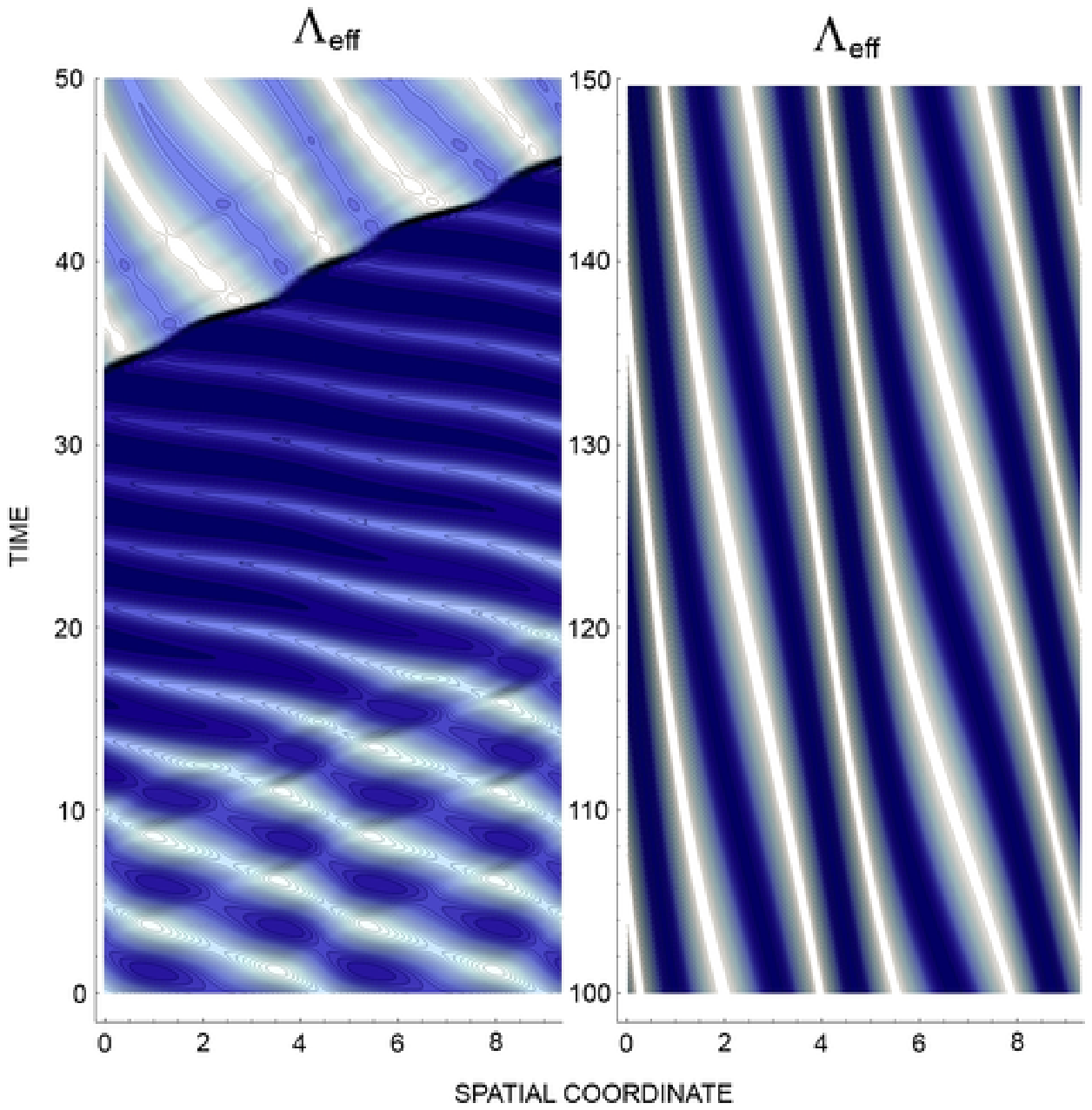}
		\caption{The vertical and horizontal axes represent the temporal and spatial dimensions respectively. The effective cosmological constant $\Lambda_{eff}$ is plotted (left) for times $t \in\left[0,50\right]$ across the range $0.25<\Lambda_{eff}<7.50$ (30 contours), with darker colours representing a smaller value. The effective cosmological constant has also been plotted (right) for later times $t \in\left[100,150\right]$ across the range $2.88<\Lambda_{eff}<8.10$ (30 contours).}
		\label{result3}
		\end{figure}

\begin{figure}[p]
	\centering
		\includegraphics[width=0.75\textwidth]{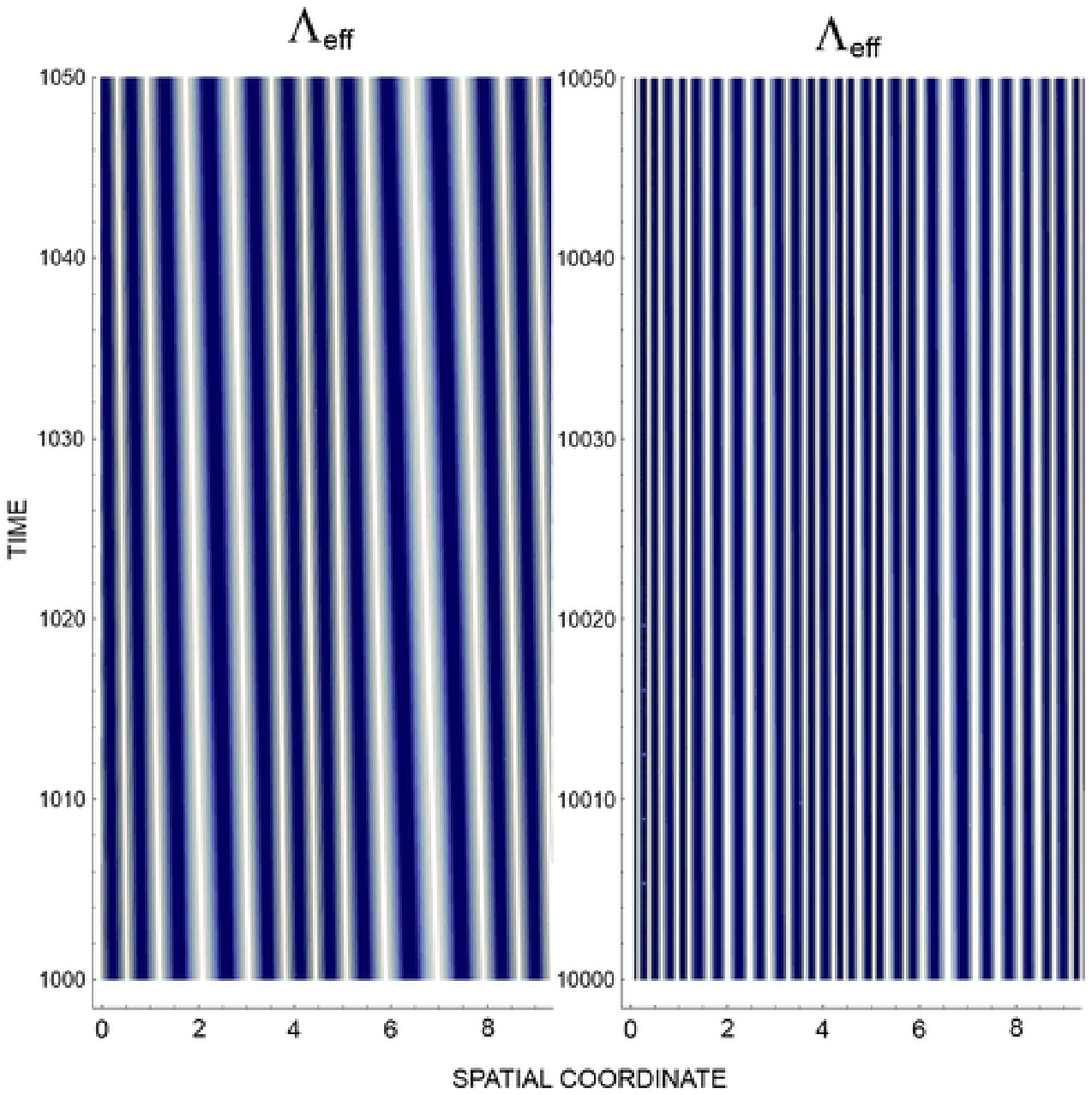}
		\caption{The vertical and horizontal axes represent the temporal and spatial dimensions respectively. The effective cosmological constant $\Lambda_{eff}$ is plotted (left) for times $t \in\left[1000,1050\right]$ across the range $3.57<\Lambda_{eff}<8.16$ (10 contours), with darker colours representing a smaller value. The effective cosmological constant has then been plotted (right) for later times $t \in\left[10000,10050\right]$ across the range $3.76<\Lambda_{eff}<7.52$ (5 contours). The key result to note is that the range of $\Lambda_{eff}$ is becoming smaller. In this present epoch, one may therefore not expect huge spatial variations within the observed cosmological constant.}
		\label{result4}
		\end{figure}

As in previous sections, the constant $\Lambda$ must be made sufficiently large, so as to prevent an infinite expansion. We also note how FIG. \ref{fig20} initially bares some resemblance to FIG. \ref{fig4}\textbf{(D)}. Except here, the scale-factor is propagating through the spatial domain, distorting the geometry of space in the process. Thus at this point, questions arise as to whether dark-energy could be the consequence of an oscillating gravitational wave. These are physical phenomena (although, as of yet, not been detected directly), which due to their localised energy-density, distort the space-time domain. This energy-density associated with the scalar field can be obtained via the effective cosmological constant of Eq.(\ref{equation15}) (cf. FIGS. \ref{result3}-\ref{result4}). 

\begin{figure}[p]
	\centering
	\includegraphics[width=0.5\textwidth]{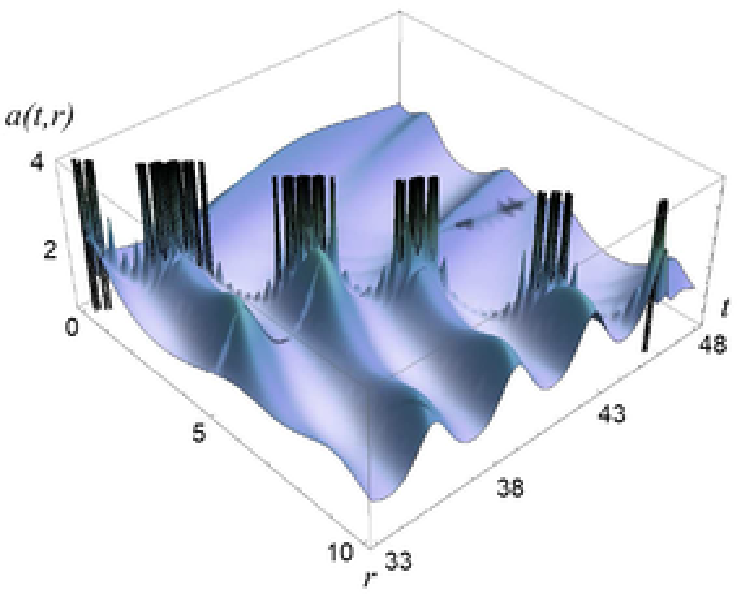}
	\caption{Space-time plot of the scale-factor $a(t,r)$. Constants are chosen as $\Lambda=3$, $\lambda=1$, $m=1$ and $\kappa=1$, subject to initial condition $a_{0}=1$. A highly localised spatial distortion can be seen moving from left to right. This is due to an interaction of the scalar field with the boundary, and propagating back along the system.}
	\label{time}	
\end{figure}

\section{A Deflationary Mechanism?}

The model which we study here, sheds light upon the various scenarios of inflation within the primordial universe \cite{Linde1982}. These are generally characterised by the choice of initial potential $V_{0}(\varphi)$ which simulates a temporarily non-vanishing cosmological term \cite{KMOS1995}. Furthermore, the classification of allowed inflationary potentials and scenarios, has been explored in \cite{KMOS1995} via the use of the catastrophe theory (cf. for details \cite{kusm}).

Inflation was first proposed more than 30 years ago now \cite{guth1981}, and suggested that our universe may have undergone a period of rapid expansion in its early stages. Three major problems exist in modern cosmology; these are dubbed the horizon, flatness, and monopole problems. The two which we are most concerned with are the flatness and horizon problems. Specifically, why does the universe appear so flat, and almost homogeneous everywhere? Of course, as the universe expands, any initial perturbations in space will be flattened out. Still, how is it that our universe can appear so homogeneous over such vast cosmological scales? For two regions of space that are not in causal contact with each other, how could such an equilibrium between the two be attained? The answer resides within inflation theory, proposing that our universe was once much smaller, with all regions residing within the causal sphere. Inflation would then expand the universe beyond all proportion, to the grandest of scales, and in a mere fraction of a second. 

Throughout FIGS. \ref{fig20}-\ref{result4}, it is evident our toy model is undergoing different phases of its evolution. At $t=0$, the scale-factor was specified as homogeneous. However, as $t$ progresses, some regions appear to undergo a cyclic expansion; the amplitude of which becoming successively larger with each oscillation. These peaks correspond to the scalar field $\varphi(t,r)$ traversing various vacuum states of the potential $V(\varphi)$. Supposing the cosmological constant $\Lambda$ were smaller in value; the scalar field would not traverse as many of these states, thus leading to less oscillations. A value similar to that used in FIG. \ref{fig4}\textbf{(A)}, would present a means of a modelling a single period of expansion.

As $t$ progresses further, the $\dot{\varphi}$ contribution begins to dominate. We also note the presence of a a highly localised spatial distortion commencing at coordinates (32,0), and ending at (46,9). This distortion is plotted within FIG. \ref{time} as a three-dimensional space-time plot. This shows a propagating singularity; the consequent effect of which, is a warping of the spatial domain. However, the process (lasting a fraction of a second) is more likened to a deflationary process - regions of space undergo momentary collapse, and then re-expanding. This phenomena propagates throughout the spatial domain with speed $v$, identical to that of the incident wave. Furthermore, the process is not global, and does not permit all regions of space to be in causal contact with one another. For such a scenario to occur, the scale-factor must be zero at all points in space at a given value of time $t$. Luckily, there is another possibility which we shall discuss momentarily. 

So how does one explain FIG. \ref{phidot} after such a process? There is a negligible $\dot{\varphi}$ contribution, yet the scale-factor $a(t,r)$ has not undergone a substantial increase. As we have noted in the previous section; when $\dot{\varphi}$ approaches zero, the scale-factor increases to a large value (cf. FIG. \ref{fig3}). To explain this, we must first consider the mechanics of the system in all their gory detail. One has a travelling wave solution, which propagates with speed $v$. When the wave solution reaches the boundary of our system, it then seeks to propagate back along the spatial axis with speed $-v$. The secondary wave carries form of a soliton, and disrupts all further incoming waves, nullifying the $\dot\varphi$ contribution. As such, there are instances when $\varphi_{r}$ either grows rapidly, or becomes zero. From Eq.(\ref{equation33}), this has clear consequences for the scale-factor. Furthermore, we note that the scalar field $\varphi$ as given by Eq.(\ref{equation32}), is not bound by the speed of light (i.e., the speed $v$ could, in principle, be greater than 1). This implies that the reflected wave from the boundary would also travel with a speed greater than light, and communicate information over vast distances. This may offer possible reprieve from the horizon problem we highlighted upon earlier.

\begin{figure}[p]
	\centering
	\includegraphics[width=0.5\textwidth]{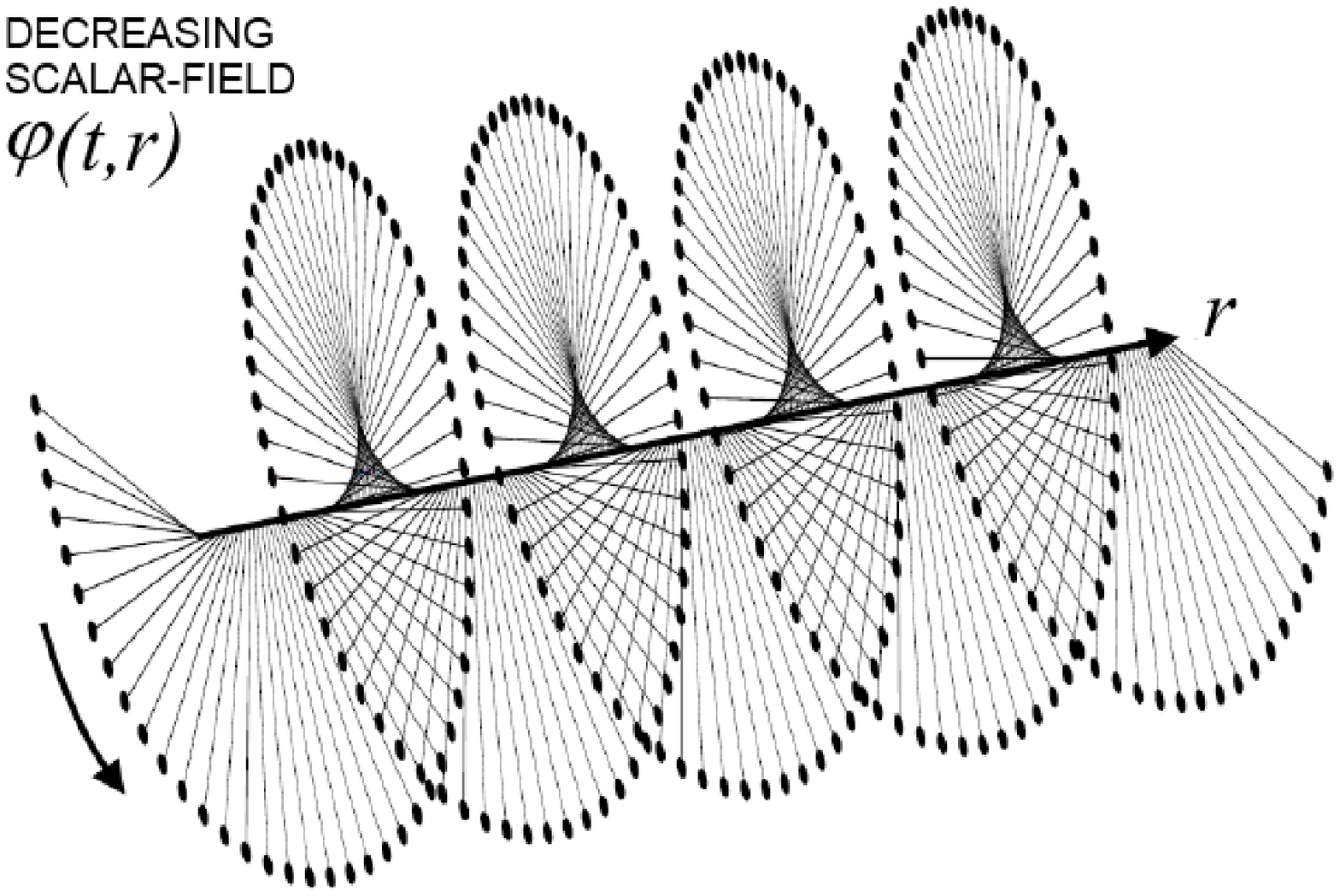}
	\caption{A physical representation of the scalar field $\varphi(t,r)$. As time $t$ progresses, the assembly performs a `screwing' motion throughout the spatial domain and travels with speed $v$. However, it is viable that the scalar gradient $\varphi_{r}$ may compete with both the scale-factor $a(t,r)$ and axion potential $V(\varphi)$, such that no screwing motion occurs. Here, the scale-factor is representative of the length for each individual pendulum. This instance first occurs at the end of the chain - the final pendulum performs a revolution, and then propagates back with speed $-v$, thus affecting all further incoming waves.}
	\label{pendula}	
\end{figure}

One must also consider the conservation of energy. With a negligible $\dot{\varphi}$ contribution, the transformed energy can only manifest itself as the scalar gradient $\varphi_{r}$, which competes with both the scale-factor $a$ and axion potential $V(\varphi)$ until equilibrium is reached. As an aside, a physical representation of the scalar field is shown in FIG. \ref{pendula}. This shows the scalar field increasing by a factor of $2\pi$ each time as $r$ increases. As time $t$ progresses, this assembly will typically perform a `screwing' motion throughout the spatial domain. 

\section{Discussion}
We have studied all possible scenarios for modelling dark-energy via the quintessence scalar field $\varphi$. Upon considering FRW models, the cosmological constant $\Lambda$ was found to be analogous to a spring-constant, and determining the elasticity of the spatial domain. As the scalar field's kinetic energy (given by the $\dot{\varphi}$ contribution) diminishes, this energy was found to transform into an elastic potential energy. This presented a desirable mechanism for expansion of the spatial domain. The consideration of an added spatial-dependence within the scale-factor also lead to a means of modelling Hubble's law. It has been found that irrespective of one's position in the universe, the more distant $r$-axis grid-lines recede at a faster rate. 

When investigating the fully inhomogeneous scenario, solutions for the scale-factor were representative of a wave-like structure, initially propagating through the spatial domain with an oscillatory amplitude. This wave had the overall effect of distorting the geometry as it travelled. For late times, the effective cosmological constant $\Lambda_{eff}$ is found to be almost homogeneous (cf. FIGS. \ref{result3} -- \ref{result4}). The variations present are characteristic of the observed microwave background, tiny fluctuations that can ultimately lead to large-scale structure formation of both filaments and voids.

A key result was also the relaxation of both the effective cosmological constant $\Lambda_{eff}$, and the effective Ricci scalar $\mathcal{R}_{eff}$, to be consistent with the small values that are observed today. From Eq(\ref{equation15}), it was found that both quantities depend upon some parameter $\xi$ which determines the scaling of some unseen extra dimension. Supposing this extra dimension $\chi$ were to be $2\pi$ periodic, the parameter $\xi$ would then determine the radius of this extra dimension. For an observed cosmological constant $\Lambda_{eff}\sim 10^{-35}$, this truly puts into perspective the energies required to access such small dimensions, well beyond that of any particle accelerator. For convenience, simulation results for FIGS. \ref{result3} -- \ref{result4} use a value $\xi=2\kappa=2$. However, this value of $\xi$ merely affects the amplitude of the result, and can be scaled as necessary.

At this stage, the cosmological constant looks improbable as an overall contribution to $\Lambda_{eff}$ (cf. Eq.(\ref{equation15})). As mentioned previously, theoretical predictions for the energy-density $\Lambda$ at the instant of the big-bang \cite{steinhardt} are of the order $\sim 10^{93}g/cm^{3}$. Using Eq.(\ref{equation15}), to relax this parameter to its present day value of $\approx 7\times 10^{-30}g/cm^{3}$, one would require an extra dimension of radius $\sim10^{-246}$m. Ideally then, this energy-density $\Lambda$ should be screened completely by the scalar field. What remains, is the tiny contribution from electromagnetism. Electromagnetic phenomena are seldom considered in gravitational physics and cosmology. However, this stark conclusion implies that their effects could be fundamental to understanding why indeed the cosmological constant is so small!

\end{document}